\documentclass[3p,11pt,times,preprint,number,sort&compress]{article}
\usepackage[utf8]{inputenc}
\usepackage{biblatex}
\usepackage{algorithm}
\usepackage{hyperref}
\usepackage{algpseudocode}
\usepackage{authblk}

\title{An effective and efficient label initialization method based on similarity for community detection in networks}

\author{Jyothimon Chandran, Madhuviswanatham Vankadara*\\ 
 \small School of Computer Science and Engg.,\\
 \small Vellore Institute of Technology, Vellore, India\\
 \small *vmadhuviswanatham@vit.ac.in
 }
\date{December 04, 2021}
\begin{document}
\maketitle

\begin{abstract}
Identifying clusters or community structures in networks has become an integral part of social network analysis. Though many methods were proposed, the label propagation algorithm (LPA) is a popular computationally efficient method with running time linear. However, the LPA provides different combination of communities on the same network due to the randomness in LPA. Many improvements have been proposed to tackle this stability problem by eliminating the randomness. This paper put forward an improvement to the standard LPA by proposing a label initialization method based on link similarity. The similarity is measured based on the connection strength between two nodes. The method is tested on real and synthetic measures to analyze the performance.
\end{abstract}

\section{Introduction}
Complex network representation of complex systems \cite{1,2} are wide spread in various domains of research including biology, computer science, and physics. It simplifies the complexity inherent in the system by denoting the components and their relations or interactions. Generally, the nodes denote the components and links indicate the interactions or relations between entities in complex networks. Community structures \cite{3} is an important property of complex networks, which helps to provide more meaningful structures in network. Understanding these structures bring insights into various dynamics of various systems \cite{1}.

There exist a plenty of literature for finding community structures in networks using various techniques \cite{4}. Some studies treat community detection as an optimization problem, in which maximizing the optimization function such as modularity are among wide spread. Since many of these methods are not computationally efficient, methods based on some heuristic rules are also recommended. Among that, label propagation algorithm \cite{5} is one of the most time efficient heuristic methods with time complexity nearly linear $(O(t.n))$ where $n$ denotes the number of nodes. However, when LPA is applied on a network, it returns community structures which are different in different runs. This shows that the LPA is unstable.

Though there are several reasons for the unstable results. The major reason is the randomness in node label update as well as the random node order. When the random labeling strategy is performed on each node, the random selection can lead to various combinations of communities. Recently, various studies propose several methods for improving LPA by recommending various label update strategies focusing on eliminating randomness. These methods focus mainly on the node label update strategy which selects labels according to an order-based selection strategy that includes node importance, label influence, and link importance.

The first step in the original LPA is to assign unique labels to every node in the network. The impact of initial node labels on the stability is not fully explored. This paper proposes a method (ILI-LPA) which initializes node labels based on node influence. The ILI-LPA eliminates the unique label initialization of LPA with a new label initialization which assigns identical labels to nodes which express high probability to be in the same community at the end the label propagation. The main aim of the paper is to assign structurally densely connected nodes with same labels as initial labels to improve the stability and accuracy of the communities. 

\section{Related works}
To eliminate random label update process in the LPA, \cite{6} proposed a method which applies modularity to avoid random selection of labels. Recently, Le et al. \cite{7} put forward a modified LPA which also applies modularity. 

Node influence is also considered as an effective method for label update to improve stability. Several methods that improve LPA on the stability of communities based on node influence are also available \cite{8,9,10}. Estimating the importance or influence of nodes is the main step in those methods. The k-shell value of each node is estimated to find the node importance in \cite{8}. Whereas in \cite{9} the node degree as well as the neighbors influence is measured to estimate node importance.  However, the increase in time complexity is a major drawback of these methods. Some studies focus on the link importance assuming that links contribute significantly to the community structure identification. The label influence is considered in \cite{11}. LPAc \cite{12} based on edge clustering coefficient is a popular method that considers the link importance through edge clustering coefficient. Similarity \cite{13} also proposed a link strength for LPA. Motifs in network are also considered for improving the LPA stability \cite{14}. Also, methods based on genetic algorithms \cite{15} and considering attributes \cite{16} are introduced. Still, the stability and accuracy is a problem unsolved.  

\section{Proposed method}
This paper recommends a node labeling method based on link similarity for finding communities in networks. The main aim of the proposed method is to improve the stability of the standard label propagation algorithm. The proposed method replaces the unique label initialization of the LPA and put forward a more accurate label initialization to minimize the impact of randomness in LPA.

The method is called ILI-LPA (Identical Label Initialization LPA). The ILI-LPA focuses on the initial node labels. The authors of this paper assume that the strategy of assigning unique labels to nodes as initial node labels has several drawbacks which affect the stability at the label propagation. This is because if unique labels are assigned to nodes in a community, the boundary nodes cannot differentiate its own community participation hence randomness create uncertainty. 

It is well known that nodes with dense connections will stay in a single community once the community detection process is over. If we can identify some of the densely connected regions and assign same labels, we can minimize the effect of randomness. Based on this concept, this paper introduces $tsi$ which evaluates the strength of connections between nodes.
\begin{equation}
    tsi(i,j) = \frac{1+|N(i) \cap N(j)|}{|N(j)|}
\end{equation}
where $N(i)$ indicates the neighbor set of node $i$. Consider node $(i,j)$ are adjacent nodes in $G ((i,j) \in E)$. The more the node $i$ is connected with the neighbors of $j$, the higher the influence of node $i$ to node $j$. The higher influence indicates not only the strength of connections, but also the probability that how much a node accept node the other. This is more meaningful to the label propagation process in LPA.
We replace the unique label assigning method of LPA to set initial node labels with identical label initialization. It is based on the following condition.
\begin{equation}
tsi(i,j) \geq tsi(j,i) \&\& tsi(i,j) \geq \beta
\end{equation}

\begin{algorithm}[H]
\caption{Identical Label Initialization-LPA }
\label{alg:the_alg}
\hspace*{\algorithmicindent} \textbf{Input:} {Network $G = (V, E)$, parameter $\beta$
}\\
\hspace*{\algorithmicindent} \textbf{Output:} {Communities $C_{D}$ = \{$C_{D1}$, $C_{D2}$, $C_{D3}$,\dots $C_{Dk}$\}}
\begin{algorithmic}[1]

\Procedure{ILI-LPA}{G,$\beta$} 
    \For {each $(i, j) \in E$ }
        \State $tsi(i,j) \leftarrow \frac{1+|N(i) \cap N(j)|}{|N(j)|}$ 
        \State $tsi(j,i) \leftarrow \frac{1+|N(i) \cap N(j)|}{|N(i)|}$ 
        \State $Weight(i,j) \leftarrow tsi(i,j)$
        \State $Weight(j,i) \leftarrow tsi(j,i)$
    \EndFor
    \For {each $i \in V$}
            \If{ $ tsi(j,i) \leq tsi(i,j) $ and $tsi(i, j) \geq \beta$}
                \State $label(j) \leftarrow label(i)$    
            \EndIf
      
    \EndFor
    \For {each $i \in V$} 
        \If {$size(l) = 1$}
            \State $label(i) \leftarrow l$
        \Else 
            \State $label(i) \leftarrow random\_select(l)$
        \EndIf
    \EndFor
    \State Repeat steps 13-19, until all node labels are similar to its neighbors majority.
    \State According to the node labels, seperate the communities

\EndProcedure
\end{algorithmic}
\end{algorithm}
Each node $i$ update the label of neighbor $j$, if the condition is satisfied. The $\beta$ indicates the input parameter. This ensures that some percent of nodes possess labels which are same, such as the nodes in densely connected regions. The original label propagation is performed after initialization of node labels. During the label propagation, each node accepts the label of its neighbors majority1y label. When more than one label satisfies majority label,  the random label selection is adopted.

The time complexity of the proposed method is reported as $O(n.d)$ where d denotes the averge degree.

\section{Results}
The algorithm is coded in Python NetworkX and experimented on an Intel i7 machine of 16GB RAM. The input parameter is assigned as value 0.35. Two quality measures, modularity \cite{22} and Normalized mutual information [nmi] \cite{23} is used to assess the performance. 

The obtained modularity on networks karate \cite{17}, dolphin \cite{18}, football \cite{3}, and polbooks \cite{19} are 0.371, 0.523, 0.604, and 0.526 respectively.  However, the modularity of LPA on the same networks are 0.357, 0.487, 0.589, 0.511 respectively. On network science \cite{20} is 0.921 and email\_enron \cite{20} is 0.562. The nmi reported in karate, dolphin, football, and polbooks are 0.837, 0.366, 0.927, and 0.593 respectively. Whereas, the LPA returns nmi values 0.649, 0.540, 0.893, 0.524. This shows that the initial node labels affect the quality of the communities.

Compared to the LPA, the ILI-LPA perform better on LFR network. On a network of 1000 nodes, the LPA modularity drops to 0 when $\mu$ is 0.55, whereas the ILI-LPA maintains quality upto  $\mu$ is 0.70. Similar pattern has been observed on LFR networks of size 2000 and 5000. This shows that the initial node labels has significant impact of the performance of the random label propagation process. 
\section{ Conclusion}
The LPA is an efficient community detection method. The stability of communities extracted by LPA is a major concern. To improve the stability, this paper proposes a label initialization method, which replaces the unique label initialization method adopted by LPA and other improvements. The proposed method identify densely connected regions based on link similarity and assigns same labels as initial node labels without consuming much computational time. The performance is verified on real-world and LFR networks. The results show effectiveness of label initialization. The main findings of the paper is that appropriate label initialization improves performance compared to distinct label initialization.


\begin{thebibliography}{99}

\bibitem{1} Newman, M. E.: The structure and function of complex networks. SIAM review, Vol. 45, 2003, No. 2, pp. 167-256.

\bibitem{2} Newman, M.: Networks. Oxford university press, 2018.

\bibitem{3} Girvan, M., Newman, M. E.: Community structure in social and biological networks. Proceedings of the national academy of sciences, Vol. 99, 2002, No. 12, pp. 7821-7826.

\bibitem{4} Xu, G., Guo, J., Yang, P.: TNS-LPA: An Improved Label Propagation Algorithm for Community Detection Based on Two-Level Neighbourhood Similarity. IEEE Access, Vol. 9, 2020, pp. 23526-23536.

\bibitem{5} Raghavan, U. N., Albert, R., Kumara, S.: Near linear time algorithm to detect community structures in large-scale networks. Physical review E, Vol. 76, 2007, No. 3, p. 036106.

\bibitem{6} Barber, M. J., Clark, J. W.: Detecting network communities by propagating labels under constraints. Physical review E, Vol. 80, 2009, No. 2, p. 026129.

\bibitem{7} Le, B. D., Shen, H., Nguyen, H., Falkner, N.: Improved network community detection using meta-heuristic based label propagation. Applied Intelligence, Vol. 49, 2019, No. 4, pp. 1451-1466,


\bibitem{8} Xing, Y., Meng, F., Zhou, Y., Zhu, M., Shi, M., Sun, G.: A node influence based label propagation algorithm for community detection in networks. The Scientific World Journal, Vol. 2014, 2014.

\bibitem{9} Zhang, X. K., Ren, J., Song, C., Jia, J., Zhang, Q.: Label propagation algorithm for community detection based on node importance and label influence. Physics Letters A, Vol. 381, 2017, No. 33, pp. 2691-2698.


\bibitem{10} Wang, T., Chen, S., Wang, X., Wang, J.: Label propagation algorithm based on node importance. Physica A: Statistical Mechanics and its Applications, Vol. 551, 2020, pp. 124137.

\bibitem{25} Zhang, Y., Liu, Y., Li, Q., Jin, R., Wen, C.: Lilpa: A label importance based label propagation algorithm for community detection with application to core drug discovery.  Neurocomputing, Vol. 413, 2020, No. 1, pp. 107-133.


\bibitem{28} Zhang, X. K., Tian, X., Li, Y. N., Song, C.: Label propagation algorithm based on edge clustering coefficient for community detection in complex networks. International Journal of Modern Physics B, Vol. 28, 2014, No. 30, p. 1450216.

\bibitem{29} Berahmand, K., Bouyer, A.: A link-based similarity for improving community detection based on label propagation algorithm. Journal of Systems Science and Complexity, Vol. 32, 2019, No. 3, pp. 737-758.

\bibitem{32} Li, P. Z., Huang, L., Wang, C. D., Lai, J. H., Huang, D.: Community detection by motif-aware label propagation. ACM Transactions on Knowledge Discovery from Data (TKDD), Vol. 14, 2020, No. 2, pp. 1-19.

\bibitem{33} Hosseini, R., Rezvanian, A.: AntLP: ant-based label propagation algorithm for community detection in social networks, CAAI Transactions on Intelligence Technology, Vol. 5, 2020, No. 1, pp. 34-41.

\bibitem{34} Berahmand, K., Haghani, S., Rostami, M., Li, Y.: A new attributed graph clustering by using label propagation in complex networks. Journal of King Saud University-Computer and Information Sciences, 2020.





\bibitem{37} Zachary, W. W.: An information flow model for conflict and fission in small groups. Journal of anthropological research, Vol. 33, 1977, No. 4, pp. 452-473.

\bibitem{38} Lusseau, D., Schneider, K., Boisseau, O. J., Haase, P., Slooten, E., Dawson, S. M.: The bottlenose dolphin community of Doubtful Sound features a large proportion of long-lasting associations. Behavioral Ecology and Sociobiology, Vol. 54, 2003, No. 4, pp. 396-405.

\bibitem{39} Krebs, V.: A network of co-purcheased books about US politics. October, Vol. 20, 2008, No. 1, pp. 0-03. 

\bibitem{41} Newman, M. E. J.: Network data, 2013, \href{http://www-personal. umich. edu/~ mejn/netdata}{www-personal. umich. edu/~ mejn/netdata}.


\bibitem{43} Lancichinetti, A., Fortunato, S., Radicchi, F.: Benchmark graphs for testing community detection algorithms. Physical review E, Vol. 78, 2008, No. 4, p. 046110.

\bibitem{44} Newman, M. E.: Modularity and community structure in networks. Proceedings of the national academy of sciences, Vol. 103, 2006, No. 23, pp. 8577-8582.

\bibitem{45} Danon, L., Diaz-Guilera, A., Duch, J., Arenas, A.: Comparing community structure identification. Journal of statistical mechanics: Theory and experiment, Vol. 2005, 2005, No. 09, p. 09008.

\end{thebibliography}
\end{document}